\documentclass[aps,twocolumn,epsf,floatfix,pre]{revtex4-1}
\usepackage{graphics,graphicx,epsfig}
\usepackage{amssymb,color}
\usepackage{epsf,epstopdf,wrapfig}
\usepackage{amsmath}
\usepackage{multirow}

\begin{document}

\title{ Error control in the set-up of stereo camera systems for 3d animal tracking}

\author{Andrea Cavagna$^{*}$, Chiara Creato$^{*,\ddagger}$, Lorenzo Del Castello$^{*,\ddagger}$, Irene Giardina$^{*,\ddagger}$,  Stefania Melillo$^{*,\ddagger}$, Leonardo Parisi$^{*,\S}$, Massimiliano Viale$^{*,\ddagger}$}

\affiliation{$^*$ Istituto Sistemi Complessi, Consiglio Nazionale delle Ricerche, UOS Sapienza, 00185 Rome, Italy}
\affiliation{$^\ddagger$ Dipartimento di Fisica, Universit\`a\ Sapienza, 00185 Rome, Italy}
\affiliation{$^\S$ Dipartimento di Informatica, Universit\`a\ Sapienza, 00198 Rome, Italy}

\begin{abstract}
Three-dimensional tracking of animal systems is the key to the comprehension of collective behavior.
Experimental data collected via a stereo camera system allow the reconstruction of the 3d trajectories
of each individual in the group. Trajectories can then be used to compute some quantities of interest
to better understand collective motion, such as velocities, distances between individuals and correlation
functions. The reliability of the retrieved trajectories is strictly related to the accuracy of the 3d reconstruction.
In this paper, we perform a careful analysis of the most significant errors affecting 3d reconstruction,
showing how the accuracy depends on the camera system set-up and on the precision of the calibration parameters.
\vspace{0.5 cm} 
\end{abstract}
\maketitle
\section*{Introduction}
Technological improvements in the field of digital cameras are strongly simplifying the study of collective behavior in animal groups. The use of single or multi-camera systems to record the time evolution of a group is by far the most common tool to understand collective motion. The emergence of collective behavior in human crowds \cite{gallup2012, moussaid2009}, fish schools \cite{katz2011, butail20103d}, bird flocks \cite{attanasi2014information, attanasi2014emergence} and insect swarms \cite{attanasi2014prl, attanasi2014collective, butail20113d, butail2012reconstructing} have been investigated. Events of interest are recorded from one or more cameras and images are then processed to reconstruct the trajectory of each individual in the group. Positions of single individuals at each instant of time are used to characterize the system. Density, mean velocity, mean acceleration, size of the group, as well as single velocities and accelerations are computed to understand how collective behavior arises and following which rules. The reliability of these results is strictly connected to the accuracy of the reconstructed trajectories.

In a previous work Cavagna et al. in \cite{cavagnaAnimal} suggested how to reduce the error on the retrieved trajectories choosing the proper set up for the system. More recently in \cite{theriault2014}, a tool to check the accuracy of a multicamera system on the reconstructed trajectories is provided, together with a software for the calibration of the intrinsic and extrinsic parameters. In \cite{towne2012}, Towne et al. give a way to measure the reconstruction error and to quantify it when a DLT technique is used to calibrate the extrinsic parameters of the system. But a theoretical discussion on  the propagation of the errors from experimental measures to the reconstructed positions and trajectories is still missing.

From a collective behavior perspective, distances between targets are much more interesting than absolute positions. Indeed, quantities like density, velocity, acceleration are not referred to the position of single animals, but they involve a measure of a distance. For this reason, in our discussion we will show how to estimate the error on the reconstructed $3d$ position of a single target, but we will give more emphasis to the error on mutual distances between two targets. We will focus the analysis on  how experimental measurements and calibration uncertainty affect mutual distances between targets and we will give some suggestions on how to choose a suitable set up in order to achieve the desired and acceptable error. 

In the first section of the paper we give a description of the pinhole model, which is by far the simplest but effective approximation of a digital camera. We introduce the nomenclature used in the entire paper. Moreover we describe the mathematical relations holding between the position of a target in the three dimensional real world and the position of its image on the camera sensor. In the same section, we describe the general principles of the $3d$ reconstruction making use of systems of two pinhole cameras. In the second section we will show the error formulation for both absolute position of a single target and mutual distance between pair of targets considering at first one camera only, and then generalizing the results to the case of a camera system. In the third section  we give an interpretation of the error formulation to suggest how the reconstruction error can be reduced by properly choosing the suitable intrinsic and extrinsic parameters of the system. In the fourth and last section we consider the set up we use in the field to record starling flocks and midge swarms. We give a description on the tests we perform to check the accuracy of the retrieved trajectories. Moreover we practically show how the results of these tests are affected by experimental measurements inaccuracy.

\section{Three dimensional reconstruction: general principles}\label{section::general}

The reconstruction of the position of a target imaged by one or more camera can be intuitively address with a geometric perspective. In some special situations, the images of only one camera gives enough information to determine where the target is located in the real world. However in the general case at least two cameras are needed. 

In this section we address the geometric formulation of the $3d$ reconstruction problem for one camera and for a system of two cameras, knowing intrinsic and extrinsic parameters of the system. Intrinsic parameters fix the geometry of the camera and its lens: the position of the image center, the focal length and the distortion coefficients. We calibrate the intrinsic parameters taking into account the radial distortion up to the first order coefficient only, while we do not consider tangential distortion. Tests presented in Section~\ref{section::test} show that this is sufficient to obtain the desired accuracy in the $3d$ reconstruction. Extrinsic parameters, instead, describe the position and orientation of the camera system with respect to a world reference frame. They generally include a set of angles, fixing the orientation of the cameras and a set of length measures expressed in meters  defining the cameras positions. Depending on the experiment they can be practically measured with high precision instruments or calibrated through software tools. In the paper, we consider cameras in the pinhole approximation, which is the easiest but effective camera model. Note that the pinhole approximation does not take into account lens distortion. In the following when talking about the position of a target on the image plane, we will always refer to its coordinates already undistorted.

\begin{figure}[t!]
\includegraphics[width=1.0\columnwidth]{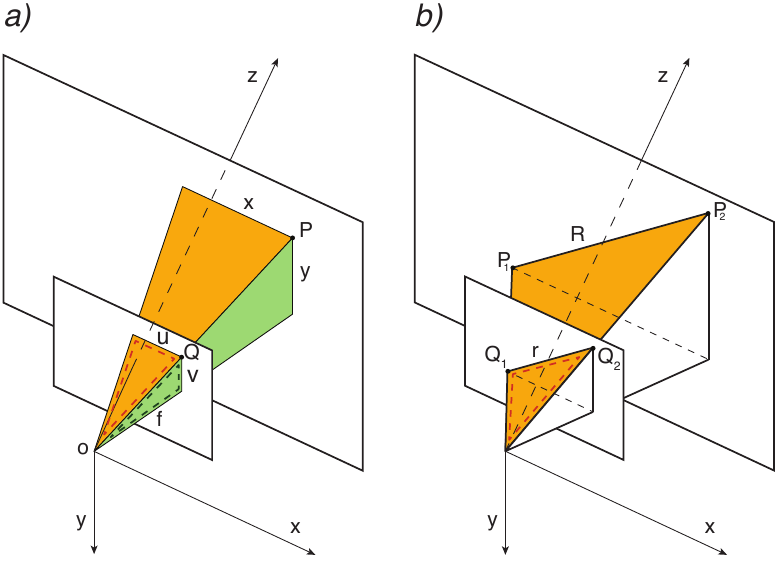}
\caption{
{\bf Pinhole model.}  
{\bf a:} Similarity between the dashed and the filled orange triangles shows the proportionality between the $x$ coordinate of a $3d$ target $P$ and the $u$ coordinate of its image $Q$ by a factor $z/\Omega$. Similarity between the two dashed and filled green triangles shows instead the proportionality between the $y$ coordinate of the point $P$ and the $v$ coordinate of its image. 
{\bf b:} The two targets $P_1$ and $P_2$ lying on the same $z$-plane are imaged in $Q_1$ and $Q_2$. Similarity between the two dashed and filled orange triangles shows that the distance $ R=|\protect\overrightarrow{P_1P}_2|$  is proportional to the $2d$ distance $r=|\protect\overrightarrow{Q_1Q}_2|$ through the coefficient $z/\Omega$.
}
\label{fig:pinhole_model}
\end{figure}

A camera maps the three dimensional real world into the two dimensional space of its image. 
 In the pinhole approximation the correspondence between the $3d$ space and the $2d$ image plane is defined as a central projection with center in the focal point $O$, see Fig.\ref{fig:pinhole_model}a. The image plane is located at a distance $f$ from the focal point $O$, with $f$ being the focal length of the lens coupled with the camera. Under the pinhole camera model, a point $P$ in the three dimensional world is mapped to the point $Q$ on the image plane belonging to the line connecting $O$ and $P$. Note that in this approximation the image plane is located in front of the focal point. Thus the resulting image on the sensor does not reproduce the scene rotated of $180^\circ$ as it would be in a real camera. We choose this representation for its simplicity and because it does not affect in any way the discussion of the present paper.

The natural $3d$ reference frame for a camera in the pinhole approximation is the one having the origin in the focal point $O$, $z$-axis coincident with the optical axis of the camera and the $xy$-plane parallel to the sensor, as shown in Fig.\ref{fig:pinhole_model}a. Coordinates in this reference frame are expressed in meters, while coordinates of a point belonging to the sensor plane are usually expressed in pixels. The size of a single pixel, $w_p\times h_p$, defines the correspondence between the two units of measurements. Thus the point $Q$ belonging to the image plane and such that $Q^\prime\equiv(x^\prime,y^\prime,f)$ in the $3d$ pinhole reference frame corresponds to $Q\equiv(u,v)$ in the sensor reference frame, with $u=x^\prime/w_p$ and $v=y^\prime/h_p$. For the sake of simplicity,  we will assume to deal with sensor made of squared pixels, i.e. $w_p=h_p$ so that the conversion factor from meters to pixel is the same in both the direction $x$ and $y$. From now on, we denote the coordinates expressed in meters by $x$, $y$ and $z$, and the coordinate expressed in pixels by $u$, $v$ while $\Omega$ represents the focal length $f$ expressed in pixels.

\subsection{Pseudo-$\mathbf{3d}$ reconstruction: single camera case}
As shown in Fig.\ref{fig:pinhole_model}b, target $P=(x,y,z)$ in the real world is projected into a point $Q=(u,v)$ on the image plane. Using similarity between triangles it can be shown that $u=\Omega x/z$ and $v=\Omega y/z$. The correspondence between the $3d$ real space and the $2d$ image plane is then defined as: 
$
(x,y,z)\rightarrow(u,v)=(\Omega x/z,\Omega y/z),
$
or equivalently:
\begin{equation}\label{eq::pinhole}
u=\Omega x/z \mbox{ and } v=\Omega y/z
\end{equation}
Thus, knowing the $3d$ position of a target we can determine the $2d$ coordinates of its image.  But in the general case we would like to do the opposite: knowing the position of a target on the sensor plane, we would like to retrieve the corresponding $3d$ coordinates in the pinhole reference frame. If no other informations are available, eqs.(\ref{eq::pinhole}) do not have a unique solution in the unknown $(x,y,z)$ so that $3d$ reconstruction is not feasible making use of one camera only, and at least two cameras are needed. 

In the special case of targets lying on a plane, extra information about the mutual position between the camera and the plane where the motion occurs, i.e. extrinsic parameters, can be used to define an homography. Eqs.(\ref{eq::pinhole}) can then be inverted and the $3d$ positions of the targets can be retrieved making use of one camera only. For the sake of simplicity, in this section we will not address this general case, but only the easier particular situation when the motion happens on a plane parallel to the camera sensor. The interested reader can retrieve the exact formulation for the general planar motion putting together the information written in this section and in the following one, where the general $3d$ case is discussed.

In the special case of targets lying on a plane parallel to the sensor, i.e. fixed $z$, eqs.(\ref{eq::pinhole}) can be inverted and the $3d$ position of a target projected in $Q=(u,v)$ can be computed as:
\begin{equation}\label{eq::pinhole_XYZ}
x=u\displaystyle\frac{z}{\Omega} \mbox{ and } y=v\displaystyle\frac{z}{\Omega}
\end{equation}


\begin{figure*}
\includegraphics[width=1.0\textwidth]{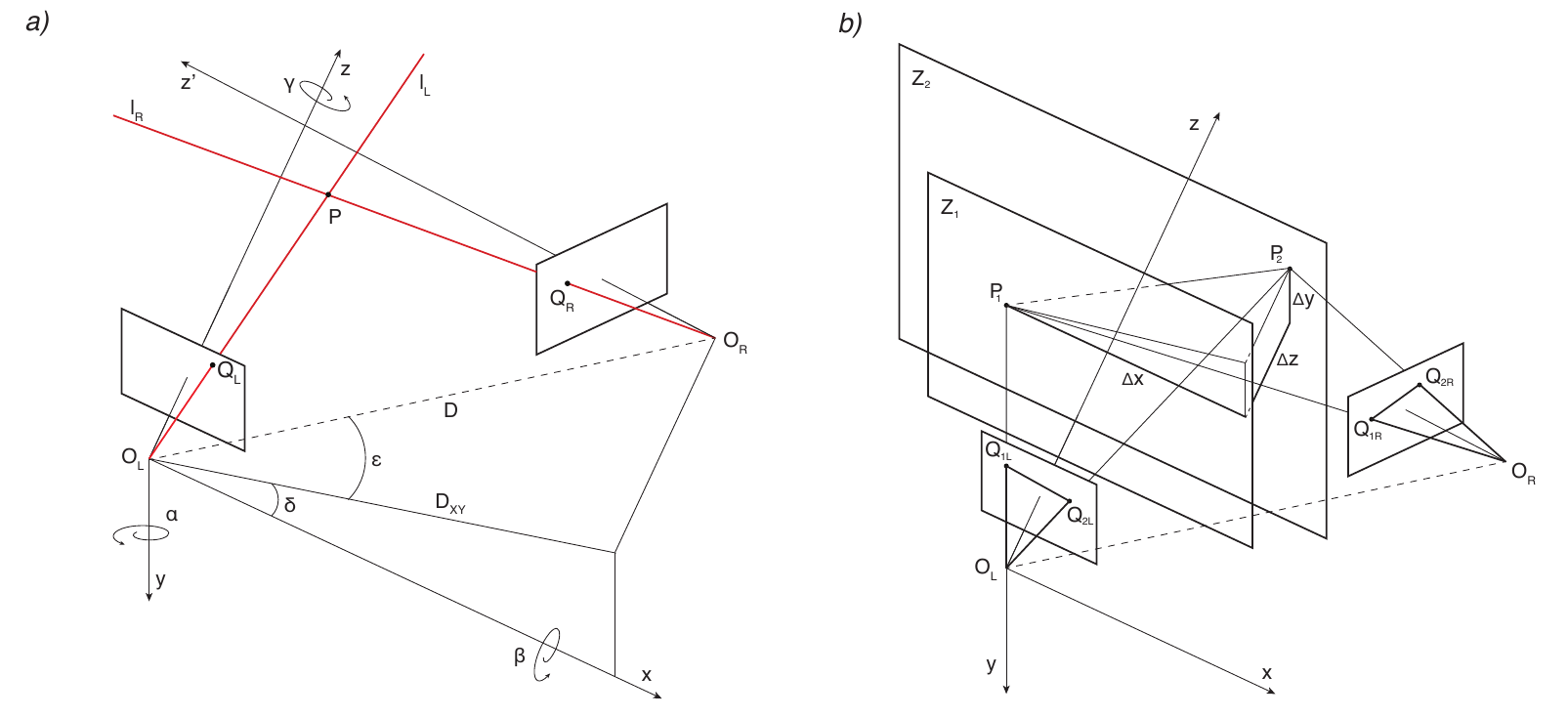}
\caption{
{\bf Pinhole camera system}.   
{\bf a: }Scheme of a two camera system. The position of a target $P$ is identified as the crossing point between the two red lines, $l_L$ and $l_R$. The mutual position of the two cameras is described by the vector $\vec{D}=d(\cos\delta\cos\epsilon, \sin\delta\cos\epsilon, \sin\epsilon)^T$ where $\delta$ is the angle between the projection of $\vec{D}$ on the $xy$-plane, $\vec{D}_{xy}$,  and the $x$-axis and $\epsilon$ is the angle between $\vec{D}$ and $\vec{D}_{xy}$. The mutual orientation between the two cameras is instead described by a rotation matrix $R$ can be parametrized through the three angles of yaw, pitch and roll, respectively denoted by $\alpha$, $\beta$ and $\gamma$.{\bf b: } The distance $ R$ between two targets $P_1$ and $P_2$ lying on different $z$ planes decomposed in its three components $\Delta x$, $\Delta y$ and $\Delta z$.
}
\label{fig:pinhole_system}
\end{figure*}

Consider now two targets $P_1\equiv(x_1,y_1,z)$ and $P_2\equiv(x_2,y_2,z)$ on the same $z$-plane, and their images $Q_1\equiv(u_1,v_1)$ and $Q_2\equiv(u_2,v_2)$ as in Fig.\ref{fig:pinhole_model}b. Thanks again to similarity between triangles, the distance between the two targets $ R$ and the distance between their projections $r$ satisfy the following equation:
\begin{equation}\label{eq::2d_deltaR}
 R= r\displaystyle\frac{z}{\Omega}
\end{equation}
so that $z/\Omega$ fixes the ratio between distances in the $3d$ and $2d$ space.

With the same argument it can be shown that:
\begin{equation}\label{eq::2d_deltaX}
\Delta x=\Delta u\displaystyle\frac{z}{\Omega} \mbox{ , }\Delta y=\Delta v\displaystyle\frac{z}{\Omega}
\end{equation}
where $\Delta x=x_1-x_2$, $\Delta y=y_1-y_2$, $\Delta u=u_1-u_2$, $\Delta v=v_1-v_2$.
 
In the field of collective behavior when we deal with cells or bacteria moving on a glass slide or fish in swallow water, the motion happens on a preferential plane. Setting up the camera in such a way that the sensor is parallel to this special plane, the displacement in the $z$ direction is negligible; $z$ can be considered constant and the equations above hold. Thus, measuring with an high precision instrument the distance, $z$,  between the sensor and the plane where the experiment takes place, it is possible to reconstruct the position of each target in the pinhole reference frame. In the following we will refer to these particular cases as $2d$ experiments.

\subsection{$\mathbf{3d}$ reconstruction: stereometric camera system}

The ambiguity of the $3d$ position of a target from its image on one camera only, can be easily solved making use of two or more cameras. For the sake of simplicity, from now on we take into account only system of two synchronized cameras, having the same focal length and sensors of the same size. 

Consider, as in Fig\ref{fig:pinhole_system}a, a three dimensional target $P$ in the $3d$ real world and its projections $Q_L$ and $Q_R$ on the sensor plane of two cameras, denoted by left and right. $P$ belongs to the line $l_L$ passing by $O_L$ and $Q_L$ as well as to the line $l_R$ passing by $O_R$ and $Q_R$. $P$ is the crossing point between the two lines $l_L$ and $l_R$ as shown in Fig.\ref{fig:pinhole_system}a.

In each of the two cameras, eq.(\ref{eq::pinhole}) holds, and the two lines $l_L$ and $l_R$ can be defined as:
\begin{itemize}
\item in the reference frame of the left camera a parametric equation, with parameter $a$ for the line $l_L$ is: $l_L=a(u_L,v_L,\Omega)$;
\item in the reference frame of the right camera a parametric equation, with parameter $b$ for the line $l_R$ is: $l_R=b(u_R,v_R,\Omega)$.
\end{itemize}
In order to find the crossing point, $P$, between $l_L$ and $l_R$ we need to express both the lines in the same reference frame: we choose the reference frame of the left camera to be the world reference frame. 

In the world reference frame $l_R=\vec{D}^T+bM(u_R,v_R,\Omega)^T$ where $\vec{D}=\overrightarrow{O_LO}_R$, $M$ is the rotation matrix which brings the world reference frame parallel to the  reference frame of the right camera, and the superscript $T$ denotes that the correspondent vector is transposed. The crossing point between the two lines is then obtained making use of the solution the system defined by $l_L=l_R$ in the unknown $a$ and $b$: 
\begin{equation}\label{eq::lLlR}
a(u_L,v_L,\Omega)^T=\vec{D}^T+bM(u_R,v_R,\Omega)^T.
\end{equation}
The solution $(a^\star, b^\star)$ identifies the position $P\equiv(x,y,z)=(a^\star u_L, a^\star v_L, a^\star\Omega)=(u_Lz/\Omega, v_Lz/\Omega, a^\star\Omega)$. Eqs. (\ref{eq::lLlR}) are well defined when $\vec{D}$ and $M$, i.e. the extrinsic parameters of the system, are known. 

$\vec{D}$ represents the vector distance between the two cameras. Its modulus $|\vec{D}|=d$ is the distance expressed in meters between the two focal points $O_L$ and $O_R$. Instead, the orientation of $\vec{D}$ can be expressed in spherical coordinates through two angles, $\delta$ and $\epsilon$: $\vec{D}=d(\cos\delta\cos\epsilon, \sin\delta\cos\epsilon, \sin\epsilon)^T$, see Fig.\ref{fig:pinhole_system}a. Denote by $\vec{D}_{xy}$ the projection of $\vec{D}$ on the $xy$-plane. $\delta$ is defined as the angle between the $x$-axis and $\vec{D}_{xy}$, while $\epsilon$ is the angle between $\vec{D}$ and $\vec{D}_{xy}$. The rotation matrix $M$ can be parametrized by the three angles of yaw, pitch and roll denoted respectively by $\alpha$, $\beta$ and $\gamma$: $M=M_{\alpha}M_{\beta}M_{\gamma}$. The mutual position and orientation of the cameras is then defined through the distance $d$ and the $5$ angles $\alpha$, $\beta$, $\gamma$, $\delta$ and $\epsilon$. These parameters are directly measured or calibrated when performing the experiment. We will show in the next section that inaccuracy on these quantities can strongly affect the retrieved position of the target $P$.

Consider, as in Fig.\ref{fig:pinhole_system}b, two targets $P_1\equiv(x_1,y_1,z_1)$ and $P_2\equiv(x_2,y_2,z_2)$ and their images $Q_{1L}=(u_{1L},v_{1L})$ and $Q_{2L}=(u_{2L},v_{2L})$ in the left camera and $Q_{1R}=(u_{1R},v_{1R})$ and $Q_{2R}=(u_{2R},v_{2R})$ in the right camera. The expression for $ R$ becomes more complicated passing from $2d$ to $3d$ experiments, since eq.(\ref{eq::2d_deltaX}) does not hold anymore. $\Delta x$ depends on both $\Delta u$ and $\Delta z$, as well as $\Delta y$ depends on $\Delta v$ and $\Delta z$. From eq.~(\ref{eq::pinhole}), $x_1=u_{1L}z_1/\Omega$ and $x_2=u_{2L}z_2/\Omega$. This implies that 
\begin{equation}\label{eq::3d_deltaX}
\Delta x=x_1-x_2=(\Delta u\bar{z}+\Delta z\bar{u})/\Omega
\end{equation} 
where $\bar{z}=(z_1+z_2)/2$, and $\bar{u}=(u_{1L}+u_{2L})/2$.  With the same argument it can be proved that $\Delta y=(\Delta v\bar{z}+\Delta z\bar{v})/\Omega$, with $\bar{v}=(v_{1L}+v_{2L})/2$. As a consequence:

\begin{equation}\label{eq::3d_deltar}
 R=\left(\displaystyle\frac{(\Delta u\bar{z}+\Delta z\bar{u})^2}{\Omega^2}+\displaystyle\frac{(\Delta v\bar{z}+\Delta z\bar{v})^2}{\Omega^2}+\Delta z^2\right)^{1/2}
\end{equation}
For short $\Delta z$, eq.(\ref{eq::3d_deltar}) becomes $ R= r\bar{z}/\Omega$ giving back eq.(\ref{eq::2d_deltaR}) for the $2d$ experiments. 

The introduction of a non constant third coordinate $z$, makes the expression of the reconstructed position not transparent. For this reason we will not discuss the general case, for further information see \cite{hartley2003book},  but in the following we will retrieve the exact solution of eq.(\ref{eq::lLlR}) for the two special cases described in Fig.\ref{fig:common_fov} and highlighted respectively in black and red.

\subsubsection{\bf{Pure translation along x axis.}}
\begin{figure}[t!]
\includegraphics[width=1.0\columnwidth]{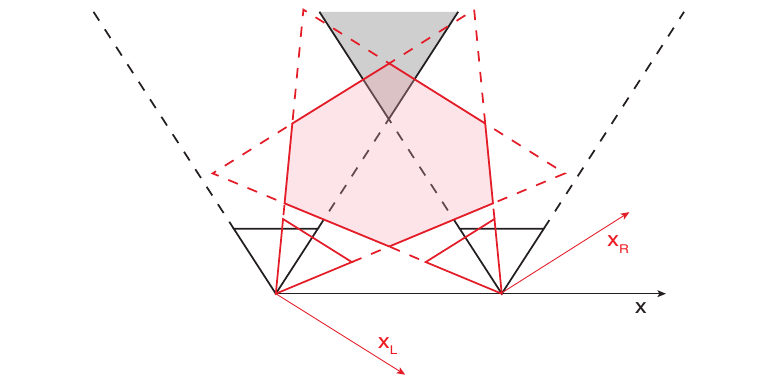}
\caption{{\bf Parallel vs rotated cameras set up.} In black a set up with the two cameras with the same orientation but with a displacement in the direction of $x$, while in red a symmetric set up with a mutual rotation about the $y$ axis only. The two different system field of views are highlighted respectively in black and red. Note that increasing the focal length of the two cameras makes the common field of view narrower. Increasing $d$ moves the working distance further in the $z$ direction and also reduces the portion of $3d$ space imaged by both cameras. Moreover the parallel set up has an optimal field of view at large $z$ while the rotated set up is optimal for short $z$, indicating that $\alpha$ affects the optimal working distance.
}
\label{fig:common_fov}
\end{figure}

In this special case, the two cameras have the same orientation and the two focal points both lie on the $x$ axis with a mutual distance equal to $d$, see Fig.\ref{fig:common_fov} where this set up is highlighted in black together with its field of view. $\vec{D}=d(1,0,0)^T$ and the rotation matrix $M$ is equal to the identity matrix. Eqs.(\ref{eq::lLlR}) become:
\begin{equation*}
a(u_L,v_L,\Omega)^T=(d,0,0)^T+b(u_R,v_R,\Omega)^T
\end{equation*}
In order to retrieve the position of the target $P$ imaged in $Q_L\equiv(u_L,v_L)$ and $Q_R\equiv(u_R,v_R)$ the above system has to be solved in the unknown $a$ and $b$. We find the solution $a^\star=b^\star=d/(u_L-u_R)=d/s$. $a^\star$ represents the ratio between the metric distance between the two focal points and the disparity $s=u_L-u_R$ expressed in pixels.

From equations~(\ref{eq::lLlR}) we obtain:
$$
\left\{
\begin{array}{rl}
x &= u_Lz/\Omega\\
y &= v_Lz/\Omega\\
z &= \Omega d/s
\end{array}
\right.
$$

\subsubsection{\bf{Translation along the x axis plus symmetric rotation about the y axis.}}
This is the special case obtained applying a translation along the $x$ axis and then rotating the left camera of an angle $\alpha/2$ in the clockwise direction and the right camera of an angle $\alpha/2$ in the counterclockwise direction about the $y$ axis, as shown in Fig.\ref{fig:common_fov} where this set up is highlighted in red.

The mutual angle of rotation about the $y$ axis between the cameras is equal to $\alpha$, so that the rotation matrix is:
$$
M=
\begin{pmatrix}
\cos\alpha & 0 & -\sin\alpha\\
0 & 1 & 0\\
\sin\alpha & 0 & \cos\alpha
\end{pmatrix}
$$

Eqs.~(\ref{eq::lLlR}) are then:

\begin{equation*}
\begin{array}{lcl}
au_L &=& d\cos\epsilon+b[u_R\cos\alpha -\Omega\sin\alpha]\\
av_L &=& bv_R\\
a\Omega &=& d\sin\epsilon+b[u_R\sin\alpha +\Omega\cos\alpha]
\end{array}
\end{equation*}

The solution of the above system is not trivial and different approximations can be made to simplify the problem. In our case we can assume that the angle of rotation $\alpha$ is small and since the set up is symmetric $\epsilon=\alpha/2$.

\begin{figure*}
\includegraphics[width=1.0\textwidth]{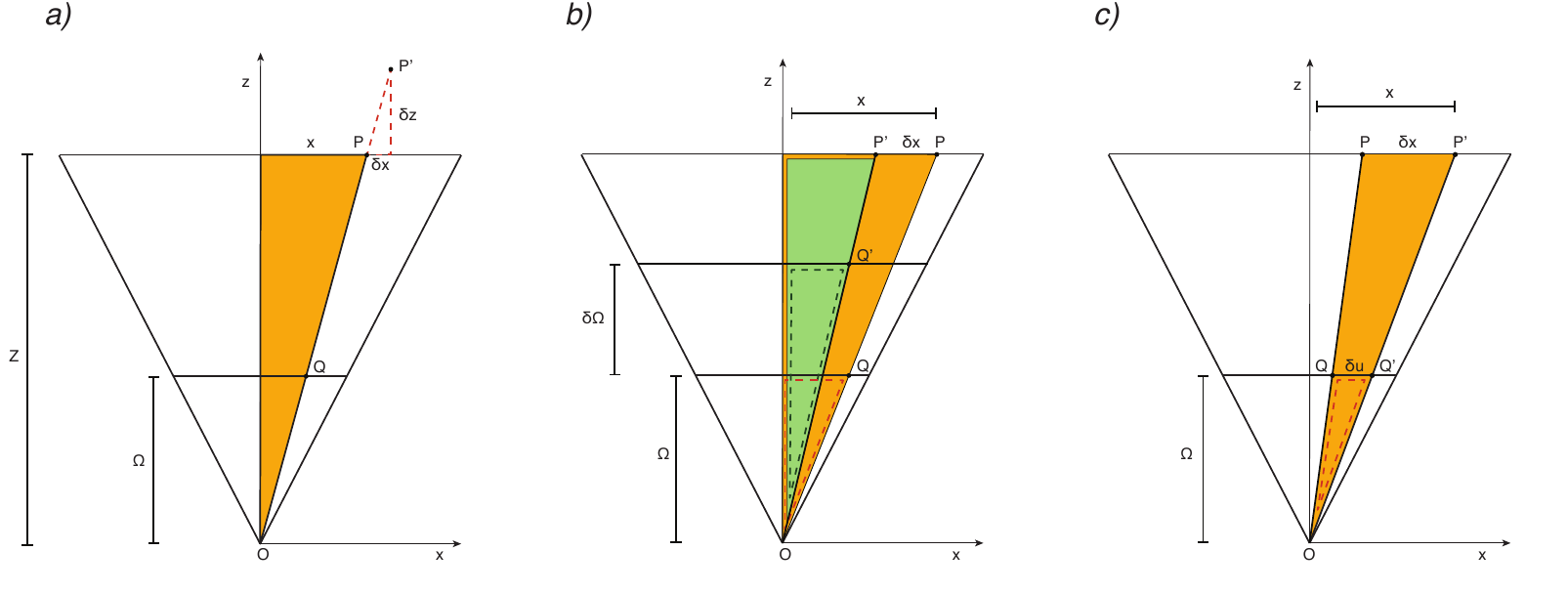}
\caption{
{\bf Errors representation for a 2d experiment}. Making use of the similarity between the filled and the dashed orange triangles and between the filled and the dashed green triangles show how the error on $z$, $\Omega$ and $u$ affect the position of a target and on the distance between a pair of targets. {\bf Panels a,b,c} refers to the error on the absolute position of $P$ due respectively to $z$, $\Omega$ and $u$. Note that in panel a) and c) $\delta x$ is positive, with $P^\prime$ lying on the right side of $P$, while in panel b) $\delta x$ is negative, with $P^\prime$ on the left side of $P$.
}
\label{fig:error_camera_system}
\end{figure*}

For small angles $\alpha$, $\sin\alpha\sim\alpha$ and $\cos\alpha\sim 1$, and the previous equations become:

\begin{equation*}
\begin{array}{lcl}
au_L &=& d+b[u_R-\alpha\Omega]\\
av_L &=& bv_R\\
a\Omega &=& d\alpha/2+b[\alpha u_R+\Omega]
\end{array}
\end{equation*}
Solving the above system we obtain:
$$
a^\star=\displaystyle\frac{1}{\Omega}\left(\Omega d - u_Ld\displaystyle\frac{\alpha}{2}\right)\displaystyle\frac{\alpha u_R+\Omega}{\alpha(u_Lu_R+\Omega^2)+\Omega s}
$$
and with the additional assumption that $u_L$, $u_R\ll\Omega$, $a^\star=d/(s+\alpha\Omega)$. So that, eqs.(\ref{eq::lLlR}) become:
\begin{equation}\label{eq::3d_xyz}
\left\{
\begin{array}{rl}
x &= u_Lz/\Omega\\
y &= v_Lz/\Omega\\
z &= a^\star\Omega=\Omega d/(s+\alpha\Omega)
\end{array}
\right.
\end{equation}
Note that for $\alpha=0rad$ the solution is exactly what we obtained in the case of pure translation.

The approximation $\sin\alpha\sim\alpha$ and $\cos\alpha\sim 1$ holds for angles approaching $0~rad$. For angles smaller than $0.2~rad$, the error in the approximation is of the third order for the sine and of the second order for the cosine, so that if $\alpha=0.2~rad$, $\sin\alpha-\alpha\sim 10^{-3}~rad$ and $\cos\alpha-1\sim 10^{-2}~rad$. When $\alpha$ is not small, eq.(\ref{eq::lLlR}) can not be simplified and the solution is not trivial anymore. For the sake of simplicity, we do not give here the formulation of the solution for the general case.

\section{Error control: theoretical relations}
In the two previous sections we described systems of one or more cameras in the pinhole approximation. We showed how to retrieve the three dimensional position of a target and the mutual distance between two targets knowing only the parameters of the system. Through the error analysis we want to quantify how errors in the experimental measures and calibration of the intrinsic and extrinsic parameters affect the reconstruction  process. Moreover we want to investigate the possibility to reduce the error choosing the proper experimental set up. We will focus our analysis on the reconstruction of the three dimensional position of a  target, but we will give more emphasis to the propagation of the error in the retrieved mutual distance between two targets. 

We will first address the error theory in the case of $2d$ experiments showing how to quantify the error making use of geometry only, then we will approach the error theory from a more formal and mathematical point of view. Finally we will consider the more general case of $3d$ experiments only in the formal way, since the geometric interpretation is not very intuitive.

\subsection{2d experiments}
This is the special case where objects move on a plane parallel to the sensor at a distance $z$ from the focal point. The position of target $P$ projected in $Q\equiv(u,v)$ is: $x=uz/\Omega$ and $y=vz/\Omega$. Instead the distance $ R$ between two targets $P_1$ and $P_2$ is computed making use of eq.(\ref{eq::2d_deltaR}); so that $R= rz/\Omega$ where $r$ is the distance in pixels between the projections of the two targets $P_1$ and $P_2$. The quantities involved are then $u,v,z,\Omega$ and we want to investigate how each of these parameters affect $x$, $y$ and $ R$.

\subsubsection{\bf Error on absolute position.} 
Consider the case when $z$ is directly measured with an error $\delta z$. Given a target $P\equiv(x,y,z)$, it is reconstructed in $P^\prime$ on the $(z+\delta z)$-plane on the line passing by $O$, $P$ and $Q$, as shown in Fig.\ref{fig:error_camera_system}a. Making use of similarity between triangles the following relation between the error on the $x$-coordinate of $P$, $\delta x$ and $\delta z$ can be shown:
\begin{equation*}
\delta x/x=\delta z/z
\end{equation*}
$\delta z/z$ is constant for all the targets in the field of view. The equation above implies that the position of each target projected on the sensor is affected by the same relative error $\delta x/x=\delta z/z$. In other words, the error $\delta x$ depends on the position $x$ of the target and the larger $x$, the larger is the error $\delta x$, while the ratio $\delta x/x$ is constant and equal to $\delta z/z$. In $2d$ experiments, the error in $z$ is the accuracy of the instrument used to measure it. If an instrument with an accuracy of $1mm$ is used on $z=10cm$, the relative error on $z$, and as a consequence of $x$, is equal to $0.01$. Note that the relative error is dimensionless. If instead the same instrument is used to take the measure of $z=10m$ the relative error becomes negligible being equal to $0.0001$, and producing a negligible relative error on $x$. While designing the set up an acceptable threshold for the relative error on $x$ has to be defined, then the working distance $z$ and the measure  instrument can be chosen accordingly.

Fig.\ref{fig:error_camera_system}b represents a system where the focal length is calibrated with an error $\delta\Omega$. This makes the sensor of the camera to be at a distance $\Omega+\delta\Omega$ from the focal point, instead of at a distance $\Omega$. 
$P$ is projected in $Q^\prime$ with the same $u$ coordinate of $Q$, but on the $(\Omega+\delta\Omega)$-plane, while the retrieved $P^\prime$ lies on the same $z$ plane of $P$ but on the line passing by $O$ and $Q^\prime$ and not on the correct one. So that its position on the $xz$-plane is $P^\prime\equiv(x+\delta x, z)$. Note that $x+\delta x<x$, meaning that the error $\delta x$ is negative, as shown in Fig.\ref{fig:error_camera_system}b. Making use of similarity between triangles, it can be shown that: $x/z=u/\Omega$ and $(x+\delta x)/z=u/(\Omega+\delta\Omega)$. Putting together these two equations we obtain:
$$
\delta x/x =-\delta\Omega/(\Omega+\delta\Omega)
$$
The negative sign in this equation indicates that a positive error on $\Omega$, produces a negative relative error $\delta x/x$, i.e. if the incorrect focal length is bigger than the correct one than the retrieved $x^\prime=x+\delta x$ is smaller than the correct $x$, as shown in Fig.{\ref{fig:error_camera_system}b}. 

As for the error on $z$, the error on $\Omega$ fixes the relative error on $x$. An error $\delta\Omega=30px$ with $\Omega=3000px$ produces a relative error on $x$ equal to $0.01$. The error on $\delta\Omega$ is generally completely due to the calibration procedure used for $\Omega$, so that it can be easily reduce using a precise calibration software.

Fig.\ref{fig:error_camera_system}c represents a system where an error $\delta u$ on the determination of the position of the projection of the target occurs. The point $P$ is then considered to be projected in $Q^\prime$ instead of $Q$.  The retrieved position $P^\prime$ lies on the same $z$-plane of $P$, but on the line passing by $O$ and $Q^\prime$. Similarity between triangles shows that:
$$
\delta x=\delta uz/\Omega
$$
The error $\delta u$ affects $x$ in a different way than the other two parameters. Unlike the error on $z$ and $\Omega$, it does not produce a constant relative error. Moreover it does not depend on the $x$ coordinate of the target. In a set up with $z=100m$ and $\Omega=3000px$, a target in the image segmented with an error $\delta u=3px$ produces an error $\delta x=0.1m$. If the position of a target is $x=1m$, an error of $0.1m$ corresponds to a relative error of $0.1$. If the error occurs on a target at $x=10m$, its retrieved position is $10.1m$ corresponding to a relative error of $0.01$. Note that, since the camera pinhole model does not include distortion effect, the error $\delta u$ includes the segmentation error due to noise on the picture and the error in the position of the point of interest when the distortion coefficient are not properly calibrated. 

The error $\delta x$  can be kept under control choosing the proper parameters of the set up, in particular the ratio $z/\Omega$. If the maximum acceptable error on $\delta x$ is defined as $c$  and $\delta u\sim 1px$, $z/\Omega$ has to be chosen in order to verify: $z\delta u/\Omega=z/\Omega<c$. 

In the general case the error on $x$ is the sum of the three contributes due to $\delta z$, $\delta\Omega$ and $\delta u$, so that:
\begin{equation}\label{eq::2d_percentage_deltax}
\delta x=x\left(\displaystyle\frac{\delta z}{z}-\displaystyle\frac{\delta\Omega}{\Omega+\delta\Omega}\right)+\delta u\displaystyle\frac{z}{\Omega}.
\end{equation}

For large $x$, targets at the edge of the field of view, the dominant term of the error is the relative part due to $\delta z$ and $\delta\Omega$, while for small $x$, targets in the center of the field of view, the dominant part is the one due to the error on $\delta u$.

Note that the entire discussion of this section could have been addressed in a more formal way simply computing the derivative of $x$  respect to the parameters $u$, $z$ and $\Omega$: $\delta x=x\left(\delta z/z-\delta\Omega/\Omega\right)+\delta uz/\Omega.$  The difference between this last equation and eq.(\ref{eq::2d_percentage_deltax}) is only in the term depending on $\delta\Omega$ but in the general case we can assume that $\delta\Omega\ll\Omega$ so that the two terms can be considered equal.

The same arguments used to retrieve the error on $x$ can be used to write the formulation of the error on the $y$ coordinate only referring the schemes in Fig.\ref{fig:error_camera_system} to the $yz$-plane:
\begin{equation}\label{eq::2d_percentage_deltay}
\delta y=y\left(\displaystyle\frac{\delta z}{z}-\displaystyle\frac{\delta\Omega}{\Omega}\right)+\delta v\displaystyle\frac{z}{\Omega}.
\end{equation}

\subsubsection{\bf Error on mutual distances between targets.}
 The error $\delta R$ on the distance $ R$ can be obtained deriving eq.(\ref{eq::2d_deltaR}) with respect to $z$, $\Omega$ and $ R$:
\begin{equation}\label{eq::2d_deltaDeltaR}
\delta R= R\left(\displaystyle\frac{\delta z}{z}-\displaystyle\frac{\delta\Omega}{\Omega}\right)+\displaystyle\frac{z}{\Omega}\delta r.
\end{equation}
On the $xz$-plane the previous equation is:
\begin{equation}\label{eq::3d_deltaDeltax}
\delta\Delta x=\Delta x\left(\displaystyle\frac{\delta z}{z}-\displaystyle\frac{\delta\Omega}{\Omega}\right)+\displaystyle\frac{z}{\Omega}\delta\Delta u.
\end{equation} 
The error on $z$ and $\Omega$ produce the same effect on the distance $\Delta x$ then on the absolute position of a target. They both induce a constant relative error on the distances between targets. The error $\delta\Delta x$ on large $\Delta x$ is higher than on small $\Delta x$. 

The third term, instead does not depend on $\Delta x$. As for $\delta x$, the first two terms of the error on $\Delta x$ can be reduced choosing a proper instrument to measure $d$ and a precise calibration software to calibrate $\Omega$, while the third term can be kept under a certain threshold choosing a set up with the proper ratio $z/\Omega$.

Referring the same arguments to the $yz$-plane: 
\begin{equation*}
\delta\Delta y=\Delta y\left(\displaystyle\frac{\delta z}{z}-\displaystyle\frac{\delta\Omega}{\Omega}\right)+\displaystyle\frac{z}{\Omega}\delta\Delta v.
\end{equation*}
Putting together the equation for $\delta\Delta x$ and $\delta\Delta y$ we find eq.(\ref{eq::2d_deltaDeltaR}).

The discussion made on $\delta\Delta x$ can be referred to $\delta R$. For large $ R$ the dominant term of the error is the constant relative error $\delta z/z-\delta\Omega/\Omega$, while for short $ R$ the dominant term is $\delta rz/\Omega$ which can be kept small choosing a set up with the proper ratio $z/\Omega$, as shown in the next section.

\subsection{3d experiments}
The error analysis is not trivial when dealing with real $3d$ experiments, i.e. targets are free to move in the entire $3d$ space without any preferential plane. The graphical interpretation of the errors is not as intuitive as in the $2d$ experiments. For this reason we find a formulation of the error on the position of a target and on distances between pairs of targets making use of derivatives. Moreover we analyze in detail only the special case introduced in the previous section: a set up with the two cameras translated on the $x$ axis and symmetrically rotated of an angle $\alpha/2$ about the $y$ axis, as shown in red in Fig.\ref{fig:common_fov}. The expression of the error in the case of a set up with parallel cameras can then be obtained imposing $\alpha=0rad$.

\subsubsection{\bf Error on absolute position.}
Under the additional hypotheses that $\alpha$ is a small angle, and $u_L$, $u_R\ll\Omega$ 
eq.(\ref{eq::3d_xyz}) holds and the position of a target $P$ projected in $Q_L\equiv(u_L,v_L)$ and $Q_R\equiv(u_R,v_R)$ in the left and in the right camera is defined by: $P\equiv(x,y,z)=(u_Lz/\Omega,v_Lz/\Omega,\Omega d/(s+\alpha\Omega))$.

$x$ and $y$ strictly depend on $z$, as well as $\delta x$ and $\delta y$ are affected by $\delta z$. For this reason in the analysis of the error on the absolute position of the targets we will focus first on the error on $z$ and then we will write the expression for $\delta x$ and $\delta y$ too.  Computing the derivative of $z$ defined  in eq.(\ref{eq::3d_xyz}), we find:
\begin{equation*}
\delta z=z\displaystyle\frac{\delta d}{d}+z\displaystyle\frac{\delta\Omega}{\Omega}\left(1-\displaystyle\frac{z}{d}\alpha\right)-\displaystyle\frac{z^2}{d}\left(\displaystyle\frac{\delta s}{\Omega}+\delta\alpha\right)
\end{equation*}
Note that negative signs in the previous equations indicate that a positive error $s$ and $\alpha$ produce a negative error on $z$. 

The relative error on $z$ is then:
\begin{equation}\label{eq::3d_percentage_deltaz}
\displaystyle\frac{\delta z}{z}=\displaystyle\frac{\delta d}{d}+\displaystyle\frac{\delta \Omega}{\Omega}\left(1-\displaystyle\frac{z}{d}\alpha\right)-\displaystyle\frac{z}{d}\left(\displaystyle\frac{\delta s}{\Omega}+\delta\alpha\right)
\end{equation}
where:
$\delta d/d$ is the relative error on the measured baseline, $\delta\Omega/\Omega$ is the relative error on the calibrated focal length, $\delta\alpha$ is the error on the measure of the angle $\alpha$ and $\delta s$ is the error on the disparity $s=u_L-u_R$. $\delta s$ represents the difference between the error in the determination of $u_L$ and $u_R$. As for the $2d$ case, an error on $s$ can be due to noise in the image but also to an error in the calibration of the distortion coefficients.

The relative error on $z$ is then made by one constant term, $\delta d/d$, and by three terms which grow linearly in $z$. The constant term due to the error on the measure of the baseline can be reduced choosing the proper instrument, as already discussed in the previous section about the error on $z$.

The other three terms, instead, can be reduced choosing the system parameters, $z$, $\Omega$ and $d$ in the proper way. A typical working distance $z$ is generally chosen and typical $\delta\alpha$ and $\delta s$ are estimated. The three linear terms of the equation above can then be kept smaller than a certain threshold, $c$, imposing the following inequalities: $z\delta s/\Omega d<c$ and $z\delta\alpha /d<c$. These two relations fix a lower bound for $\Omega$ and for $d$. 

Concerning $x$, substituting eq.(\ref{eq::3d_percentage_deltaz}) in eq.(\ref{eq::2d_percentage_deltax}) we find that:
$$
\delta x=x\left[\displaystyle\frac{\delta d}{d}-\displaystyle\frac{z}{d}\left(\alpha\displaystyle\frac{\delta \Omega}{\Omega}+\displaystyle\frac{\delta s}{\Omega}+\delta\alpha\right)\right]+\delta u\displaystyle\frac{z}{\Omega}
$$
and  substituting eq.(\ref{eq::3d_percentage_deltaz}) in eq.(\ref{eq::2d_percentage_deltay}):
$$
\delta y=y\left[\displaystyle\frac{\delta d}{d}-\displaystyle\frac{z}{d}\left(\alpha\displaystyle\frac{\delta \Omega}{\Omega}+\displaystyle\frac{\delta s}{\Omega}+\delta\alpha\right)\right]+\delta v\displaystyle\frac{z}{\Omega}
$$
Note that an error on $\Omega$ does not affect the three components $x$, $y$ and $z$ in the same way. In the expression for $\delta z/z$, the coefficient of the term due to $\delta\Omega/\Omega$ is equal to $(1-\alpha z/d)$ , while for $\delta x/x$ and $\delta y/y$ it is equal to $-\alpha z/d$. A positive error on $\Omega$ produces a negative error on $x$ and $y$, while the error on $z$ can be positive or negative, depending on the position of the target.

When the angle $\alpha$ is not small the approximation $\sin\alpha\sim\alpha$ does not hold anymore and the original system of equations has to be used. In the general case the complete expression of $z$ has to be derived with respect to all the variables and the expression of the error gets much more complicated, including extra terms. The sources of errors are always the same, i.e. measure or calibration errors for intrinsic and extrinsic parameters and segmentation inaccuracy, but their contributions are different. For the sake of simplicity we do not discuss here the general case. The reader interested in error formulation for the general problem has only to compute partial derivatives of the complete solution with respect to each parameter included in the expression for $z$.

\subsubsection{\bf Error on mutual distances between targets.} 
This is by far the more interesting issue. All the analysis we will do on trajectories is not based on the absolute position of the targets, but on their mutual position. In order to guarantee the accuracy we want on our analysis, we need to have accurate measure of the distances between targets. Consider two targets in the three dimensional space, $P_1$ and $P_2$, and their distance $ R=|\overrightarrow{P_1P_2}|=(\Delta x^2+\Delta y^2+\Delta z^2)^{1/2}$, as shown in Fig.\ref{fig:pinhole_system}. In the special case when $\Delta z\sim0$ the $3d$ error on the mutual distance between the two targets is essentially the case of the  $2d$ experiment and the  error on the mutual distances is $ R\sim rz/\Omega$ as shown in the previous section. Instead when $ R\sim\Delta z$ the error analysis is much more complicated. In the following we will retrieve a formulation of $\delta\Delta z$ for the special set up with a translation along the $x$ axis and a symmetric rotation about the $y$ axis of an angle $\alpha/2$. 

From eq.(\ref{eq::3d_xyz}) $z=\Omega d/(s+\alpha\Omega)$. 
$$
\Delta z=\Omega d\left[\displaystyle\frac{1}{s_1+\alpha\Omega}-\displaystyle\frac{1}{s_2+\alpha\Omega}\right]
$$
where $s_1$, $s_2$ represent the disparity of the projection of $P_1$ and $P_2$. Deriving the above equation we find:
\begin{align*}
\delta\Delta z= \Delta z&\left[\displaystyle\frac{\delta d}{d}+  \displaystyle\frac{\delta\Omega}{\Omega}\left(1-2\alpha\displaystyle\frac{\bar{z}}{d}\right)-2\displaystyle\frac{\bar{z}}{d}\left(\displaystyle\frac{\delta\bar{s}}{\Omega}+\delta\alpha\right)\right]+\\
-& 2\displaystyle\frac{\bar{z}^2}{\Omega d}\delta\Delta s
\end{align*}
where $\bar{z}=(z_1+z_2)/2$, $\delta\bar{s}=(\delta s_1+\delta s_2)/2$ and $\delta\Delta s=\delta s_1-\delta s_2$ is the difference between the error on the disparity of the two targets. 

For large $\Delta z$ the first term of the previous equation is the dominant part of the error and:
\begin{equation}\label{eq::large_Deltaz}
\displaystyle\frac{\delta\Delta z}{\Delta z}\sim \displaystyle\frac{\delta d}{d}+\displaystyle\frac{\delta\Omega}{\Omega}\left(1-2\alpha\displaystyle\frac{\bar{z}}{d}\right)-2\displaystyle\frac{\bar{z}}{d}\left(\displaystyle\frac{\delta \bar{s}}{\Omega}+\delta\alpha\right)
\end{equation}

Passing from $2d$ to $3d$ experiments the relative error on the mutual distances between two targets is not constant anymore. The only constant term is $\delta d/d$, while all the others depend linearly on the position of the two targets $P_1$ and $P_2$. The ratio $\bar{z}/d$ controls how much $\Delta z$ is affected by $\delta s$, $\delta\alpha$ and $\delta\Omega$. So that the error can be kept low choosing $z/d$ smaller than a desired value.

On the other side, for small $\Delta z$ the dominant part of the error is:
\begin{equation}\label{eq::short_Deltaz}
\delta\Delta z\sim-2\displaystyle\frac{\bar{z}^2}{\Omega d}\delta\Delta s
\end{equation}

This term is not relative, but absolute. Each pair of targets segmented with an error $\delta\Delta s$ is affected by the same error on $\Delta z$, independently on the size of $\Delta z$. Thus, this error has a bigger effect on short distances then on large ones. Moreover the dependence on $z^2$  makes the error growing very fast when the targets get farther from the cameras.  When designing the experiment it is very important to estimate this error, and to choose the set up in order to keep it small, because it will affect all the small distances. 

In the general case the two expressions in eq.(\ref{eq::large_Deltaz}) and eq.(\ref{eq::short_Deltaz})  contribute together at the error on $\Delta z$. Note that the segmentation error appears in two different terms, one linearly depending on $z$, $\Delta z \bar{z}\delta s/\Omega d$, which is due to the error in the segmentation of the single pair of targets mostly affecting large distances. The other one grows with $z^2$ and depends on the difference between the errors on the segmentation of the two pairs of targets and mostly affecting short distances. 

With similar arguments it can be shown that:
\begin{align*}
\delta\Delta x= \Delta x&\left[\displaystyle\frac{\delta d}{d}-2\displaystyle\frac{\bar{z}}{d}\left(\alpha\displaystyle\frac{\delta\Omega}{\Omega}+\displaystyle\frac{\delta\bar{s}}{\Omega}+\delta\alpha\right)\right]+2\displaystyle\frac{\bar{z}^2}{\Omega d}\delta\Delta u
\end{align*}
and 
\begin{align*}
\delta\Delta y= \Delta y&\left[\displaystyle\frac{\delta d}{d}-2\displaystyle\frac{\bar{z}}{d}\left(\alpha\displaystyle\frac{\delta\Omega}{\Omega}+\displaystyle\frac{\delta\bar{s}}{\Omega}+\delta\alpha\right)\right]+2\displaystyle\frac{\bar{z}^2}{\Omega d}\delta\Delta v
\end{align*}
The error on $ R=(\Delta x^2+\Delta y^2+\Delta z^2)^{1/2}$ is then: $\delta R=(\Delta x\delta\Delta x+\Delta y\delta\Delta y+\Delta z\delta\Delta z)/ R$
So that, for large $R$ the error, $\delta R$, is dominated by:
\begin{equation}\label{eq::3d_rel_dR}
\displaystyle\frac{\delta R}{ R}\sim \displaystyle\frac{\delta d}{d}-2\displaystyle\frac{\bar{z}}{d}\left(\alpha\displaystyle\frac{\delta\Omega}{\Omega}+\displaystyle\frac{\delta\bar{s}}{\Omega}+\delta\alpha\right)+\displaystyle\frac{\Delta z^2}{ R^2}\displaystyle\frac{\delta\Omega}{\Omega}
\end{equation}
While for short $R$ the dominant part of the error is the absolute term:
\begin{equation}\label{eq::3d_abs_dR}
\delta R\sim -2\displaystyle\frac{\bar{z}^2}{\Omega d}\delta\Delta s
\end{equation}

\section{Error control: setting up the system}
\begin{figure}[t!]
\includegraphics[width=1.0\columnwidth]{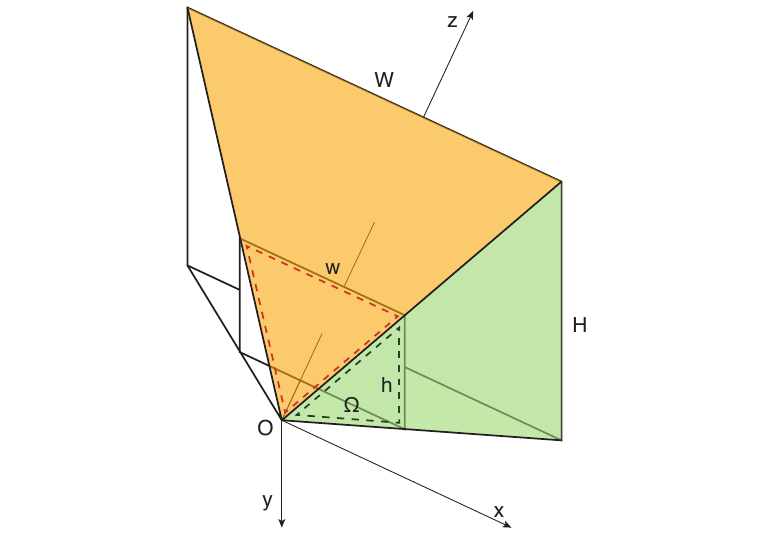}
\caption{{\bf Field of view of a single camera.} Similarity between the dashed and the filled orange triangles show that the width of the field of view, $W$, grows linearly with the sensor width, $w$, with the coefficient of proportionality equal to $z/\Omega$. While through the similarity between the dashed and the filled green triangles it can be shown that $H=hz/\Omega$.
}
\label{fig:single_fov}
\end{figure}

Designing the set up of a $3d$ experiments, intrinsic and extrinsic parameters of the system have to be chosen taking into account the volume of the $3d$ space to be imaged by the cameras,  and the accuracy of the $3d$ reconstruction. In this section we give some suggestions on how to choose the properly set up when performing $2d$ and $3d$ experiments making use of the theoretical relations for the error described in the previous sections.

\subsection{Single camera.}

When dealing with one camera only, the magnification ratio, $z/\Omega$ plays a crucial role in the choice of the set up. As shown in Section~\ref{section::general}, the magnification ratio fixes the correspondence between distances expressed in meters in the real world and distances expressed in pixels units on the sensor plane. The magnification ratio has to be chosen very carefully taking into account some properties of the objects to be tracked, but also taking care of the desired accuracy of the $3d$ reconstruction. 

First of all, an object of size $l$ in the real world would be imaged in an object of size $w_l$ on the sensor, such that $w_l=l/(z/\Omega)$. The smaller the magnification ratio, the bigger the imaged object on the screen. The previous relation can be seen as a way to fix a lower bound for $z/\Omega$. As an example, in our experience we want the image of the objects of interest to be at least as large as four pixels. Thus, when recording birds with body size $l\sim 0.4~m$, in order to have their image as large as $4$ pixels we need $z/\Omega$ to be larger than $0.1~m/px$, while when recording midges with body size of about $2~mm$, the magnification ratio should be larger than $0.0005~m/px$.

A second issue is related to the minimum appreciable distance. With the same argument used above, it can be shown that the minimum reconstructable metric distance in the $3d$ real world corresponds to one pixel on the sensor and it is defined by $r_{min}=1*z/\Omega$ and expressed in meters. This means that two objects at a mutual distance shorter than $r_{min}$ can not be distinguished in the picture. It is generally very useful to have an estimate of the interparticle distance of the group of interest and choose $z/\Omega$ in such a way that on average the distance between imaged objects is larger than $3$ or $4$ pixels. Otherwise objects would be too close to each other and optical occlusions would occur frequently. As an example if the interparticle distance is about $10cm$, we want $z/\Omega$ to be greater than $0.025m/px$.

The third aspect related to $z/\Omega$ is the choice of the size of the field of view.  
Denoting by $W$ and $H$ the width and the height of the field of view, it is easy to show that $W=wz/\Omega$ and $H=hz/\Omega$ where $w\times h$ represents the size of the sensor, see Fig.\ref{fig:single_fov}. The larger the ratio $z/\Omega$ the larger the field of view.
Denoting by $W^\star$ and $H^\star$ the minimum size of the field of view, we would like to choose $z$ and $\Omega$ such that: 
\begin{equation}\label{eq::2d_setup_zO_wh}
z/\Omega\geq W^\star/w \mbox{ and } z/\Omega\geq H^\star/h.
\end{equation}

The fourth and last issue is related to the error control. Eq(\ref{eq::2d_deltaDeltaR}) tells that the error on the distance $ R$ is: $\delta R= R(\delta z/z-\delta\Omega/\Omega)+\delta rz/\Omega$. The first two terms of this equation are constant and depend only on the precision in the measure of $z$ and in the calibration of $\Omega$, so that they can be kept as small as we want only choosing the measurement instrument with the proper accuracy. Instead the last term depends on the chosen set up and the larger the magnification ratio, the larger the absolute error on short distances.
Denote by $c$ the maximum acceptable error. Given an estimate of $\delta\Delta u$ we would like to choose $z$ and $\Omega$ such that:
\begin{equation}\label{eq::2d_setup_zO_cu}
z/\Omega\leq c/\delta\Delta u.
\end{equation}

The first three issues above give lower bound for the magnification ratio, while the last one gives an upper bound. In principle one would like to have a large field of view and a small error, but they are both controlled by $z/\Omega$, so that a compromise between the two issues has to be found. Note that the pixel size is crucial for the two problems related to the object size and the interparticle distance, while the size of the sensor plays a crucial role in the two inequalities for $W$ and $H$. In practice, the ratio $z/\Omega$ is chosen to guarantee the desired accuracy through eq.(\ref{eq::2d_setup_zO_cu}) and then the size of the sensor needed is determined by eq.(\ref{eq::2d_setup_zO_wh}).

\subsection{Two cameras system}
%
In the case of real $3d$ experiments the choice of the parameters is a bit more complicated. For the sake of simplicity, we refer only to a symmetric set up with a rotation about the $y$-axis. This is the set up we use when performing our experiment on bird flocks. It has the big advantage that the angle $\epsilon$ can be derived from the measure of the angle $\alpha$, reducing the number of experimental parameters. 

The considerations made above about the lower bound for the magnification ratio in order to guarantee the desired size of the imaged objects and the desired interparticle distance on the sensor plane, are still valid when designing a multicamera set up, but unlike $2d$ experiment, the volume of interest is not determined anymore by the field of view of one camera only. What matters now is the common field of view of the two cameras. The size of the common field of view does not depend only on $z$ and $\Omega$ but also on $d$ and $\alpha$, see Fig.\ref{fig:common_fov}. $\Omega$ influences the angle of view of each camera, so that the larger $\Omega$ the narrower each field of view and as a consequence the narrower the common field of view. $d$ affects the portion of $3d$ space in the common field of view. The larger $d$ the smaller the portion of $3d$ space imaged by the cameras. $\alpha$ affects the distance from the cameras of the common field of view. An angle $\alpha=0rad$, see Fig.\ref{fig:common_fov} where this set up is highlighted in black, makes the common field of view optimal for very large $z$. While $\alpha\neq 0 rad$ makes the common field of you optimal for short distances. In particular, the larger $\alpha$ the shorter the working distance.

The same parameter, $z$, $\Omega$, $d$ and $\alpha$ , control also the accuracy of the reconstructed distances. In fact, from eq.(\ref{eq::3d_rel_dR}) and eq.(\ref{eq::3d_abs_dR}), the error on $ R$ is
\begin{align*}
\delta R=R&\left[\displaystyle\frac{\delta d}{d}-2\displaystyle\frac{\bar{z}}{d}\left(\alpha\displaystyle\frac{\delta\Omega}{\Omega}+\displaystyle\frac{\delta\bar{s}}{\Omega}+\delta\alpha\right)+\displaystyle\frac{\Delta z^2}{ R^2}\displaystyle\frac{\delta\Omega}{\Omega}\right]+\\
&-2\displaystyle\frac{\bar{z}^2}{\Omega d}\delta\Delta s
\end{align*}

The constant term $\delta d/d$ depends only on the instrument used to take its measure and it can be strongly reduced choosing an instrument with the proper accuracy. The three terms linear in $z$ are controlled by the ratio $z/d$, while the last term by the ratio $z^2/\Omega d$. In principle the larger $d$ and $\Omega$ the lower the error, while $z$  should be as short as possible. In practice many environmental constraints are involved in the choice of the parameters and a trade off between the biological characteristic of the group of interest and the accuracy has to be found.

As for the $2d$ experiment, if we denote by $c$ the acceptable threshold for the absolute error on short $R$ and by $c^\prime$ the acceptable relative error for the large distances, we can define the set up the system imposing the two following inequalities:
$$
2\displaystyle\frac{z^2}{\Omega d}\delta\Delta s<c \mbox{ and } 2\displaystyle\frac{z}{d}\left(\displaystyle\frac{\delta\Omega}{\Omega}+\delta\alpha+\displaystyle\frac{\delta s}{\Omega}\right)< c^\prime 
$$
The above inequalities can be used to define an upper bound for both the ratios $z/d$ and $z/\Omega$ and find a set of suitable parameters which allow accuracy in the $3d$ reconstruction in the desired common field of view, respecting also the constraint due to the objects size and interparticle distances.

In many cases some of the parameters are fixed by the location where the experiments is performed. 

Indeed, when designing our experiment on bird flocks, we could not choose $z$. The experiment is performed on the roof of a building and birds are almost at $125m$ from the cameras. We can not go closer. Moreover, the baseline can not be larger than $25~m$. We put the cameras the furthest we can, so that the ratio $z/d$ is defined by the environmental constraint and it is equal to $5$. We estimated $\delta\Omega/\Omega=0.001$, $\delta\alpha=0.001~rad$ (when directly measured through the method described in \cite{cavagnaAnimal}) and $\delta s=1~px$. As a consequence $c^\prime\sim 0.01$, telling that the relative error on large distances is smaller than $0.01$. Instead, for short distances we choose $c=0.4~m$, which is a typical bird to bird distance and we estimate $\delta\Delta s\sim 0.5~px$. The previous inequality gives, than, the following lower bound for $\Omega>2z^2\delta\Delta s/c d\sim 1500$.

Instead, we perform the experiment on midge swarms in a park and  we can go as close as we want to the swarm. The working distance is than not fixed by environmental constraints. But we can not choose $d$ as large as we want. In fact, we take pictures of midges using the scattering of the sun light, so that they appear as white dots on a black background. For very large $d$ it is difficult to have a good scatter effects for both the cameras. For this reason when performing the experiments with swarms we first fix the maximum $d$ and then we choose $z$ and $\Omega$ accordingly. In practice we put the cameras the furthest possible and we set the working distance choosing $z$ in such a way to guarantee a small error on the short distances. For this experiment we choose $\Omega=7000px$, because we do not need a wide field of view since swarms are generally very stable. We define $c\sim 0.002~m$ which is the body length of a midge. The inequality above, implies that $z$ should verify: $z^2<c\Omega d/2\delta\Delta s$.  If the baseline is $6~m$, and we estimate $\delta\Delta s\sim 0.5~px$ than $z^2<84~m^2$ and we find that $z<9~m$.                                                       

\section{Error control: reconstruction tests}\label{section::test}

Every time the experiment is performed, an estimate of the reconstructed error should be taken, in order to check the experimental accuracy in measuring and calibrating that specific set up. The idea is to put some targets in the common field of view of the cameras, to measure their distance with a precise instrument, and to reconstruct their $3d$ positions. The comparison between the measured distances between pairs of targets and the reconstructed distances tells how accurate the reconstruction is. Moreover, a careful analysis of the results can reveal the source of inaccuracy and can be used when trying to fix problems.

Note that the theoretical formulation of the reconstruction problem described in this paper is meant to give an estimate of the errors when designing the experiment. In practice, eqs.(\ref{eq::lLlR}) in general does not have an exact solution. This happens because of the error in the segmented objects due to image noise and to all the errors in the measure and calibration of intrinsic and extrinsic parameters. An approximation of eqs.(\ref{eq::lLlR}) is then found, generally making use of a least squares method. 

We perform experiments in the field with starling flocks and midge swarms. The camera system set up is similar in both cases. We use two synchronized cameras shooting at $170$ fps. For flocking events we choose a baseline of $25$m and a working distance of $125$m, while for swarming events the baseline is about $6$m with a working distance of $8$m. The main difference between the two systems is the way we measure and calibrate the extrinsic parameters.

\subsection{Postcalibration: swarms}
\begin{figure}[t!]
\includegraphics[width=0.8\columnwidth]{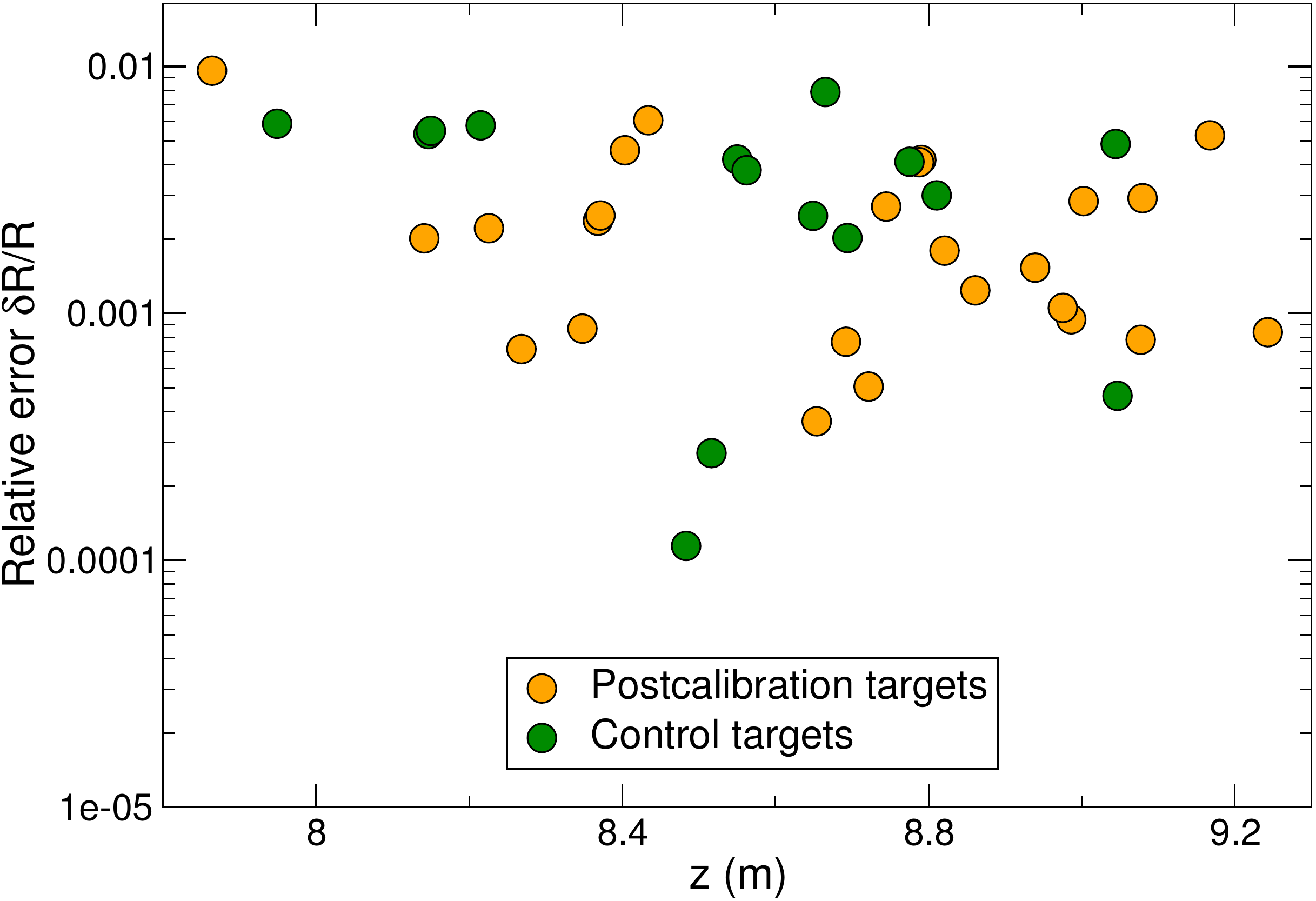}
\caption{{\bf Midges swarms. Relative error on mutual distances between targets.} Orange circles represent relative errors on the targets used in the extrinsic parameters postcalibration process, while green circles represent the relative errors on targets not used in the calibration. For both the sets of measures the relative error is lower than $0.01$.
}
\label{fig:midges_postcal}
\end{figure}

For swarming events we decide the orientation of each camera independently; we find the interesting swarm, we fix the baseline and then we rotate each camera in order to center the swarm in the image. We measure the baseline but we do not directly measure the mutual orientation of the stereometric cameras. Instead we retrieve the $5$ angles $\alpha$, $\beta$, $\gamma$, $\delta$ and $\epsilon$ making use of a post calibration procedure. Two targets, $2\times 2$ checkerboard, are mounted on a bar and their distance is accurately measured. $25$ pictures of the targets are taken in different positions, moving the bar in the $3d$ volume where the event of interest take place. A montecarlo algorithm is then used to find the $5$ angles minimizing the error in the reconstruction of the distances between the postcalibration targets. In addition we take some pictures of the targets on the bar, which are not used for the calibration procedure but only to check the reconstruction error. Typical reconstruction errors for the targets used during the calibration process are shown in Fig.\ref{fig:midges_postcal}, orange circles, and compared with the reconstruction error on the control targets not used in the calibration process, green circles. The errors on the two sets of targets are comparable and in both cases the relative errors are lower than $0.01$. This guarantees the reliability of our retrieved trajectories. 

\subsection{Precalibration: flocks}
\begin{figure}[t!]
\includegraphics[width=1.0\columnwidth]{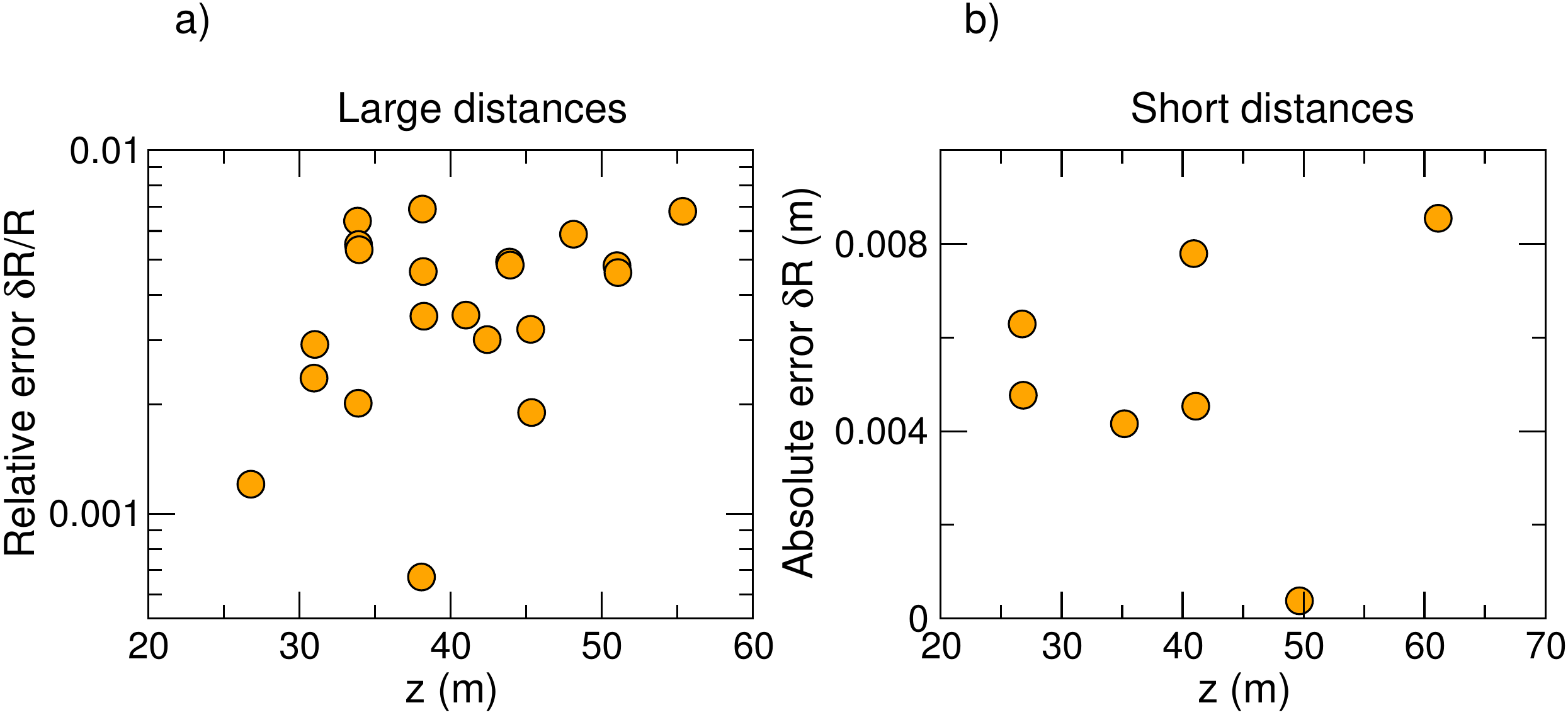}
\caption{{\bf Birds flocks. Errors on mutual distances between targets in the reconstruction test.} 
{\bf a:}  Relative errors in the reconstruction of the distance between targets at with large $ R$, $ R$ between $5m$ and $40m$. Relative errors are lower than $0.01$.
{\bf b:}  Absolute reconstruction  errors on short distances of about $0.2m$. For all the targets the absolute error is lower than $1cm$.
}
\label{fig:birds_precal}
\end{figure}

In the set up for the experiment on birds, we can not use a post calibration procedure, because we would need to take pictures of targets in the sky at at least $100$m from the cameras, nor we can take pictures of known targets to check the quality of the $3d$ reconstruction. For this reason we fix the mutual orientation of the cameras a priori as described in \cite{cavagnaAnimal}, and we record only those events happening in the common field of view. But we still need to check the accuracy. For this aim we perform reconstruction tests in a different location, setting up the cameras in a smaller set up. We want to check errors especially on the reconstruction of large distances $R$, which are the ones affected by errors in the measure of intrinsic and extrinsic parameters. For this reason we perform reconstruction tests, keeping the ratio $z/d$ as in the field. Thus we choose a baseline of $10$m and we put targets at a distance in $z$ between $20$m and $60$m. 

We accurately measure the distances between all pairs of targets. We take a picture of those targets and then we use the measured extrinsic parameters to reconstruct the distances between pairs of targets. The difference between the measured distances and the reconstructed ones gives the error on the $3d$ distances. Fig.\ref{fig:birds_precal}a shows typical relative errors for our $3d$ reconstruction test on targets at a large mutual distance $ R$. As shown in the plot, our reconstruction error is smaller than $0.01$. In Fig.\ref{fig:birds_precal}b absolute reconstruction errors on the distances of targets at short $ R\sim 0.2m$ are shown. The short distance of $0.2m$ is chosen to simulate the distance between birds in a quite dense flock. The results in Fig.\ref{fig:birds_precal}b show that we have errors of the order of $1cm$, showing the high quality of the reconstructed distances. 
\begin{figure}[t!]
\includegraphics[width=1.0\columnwidth]{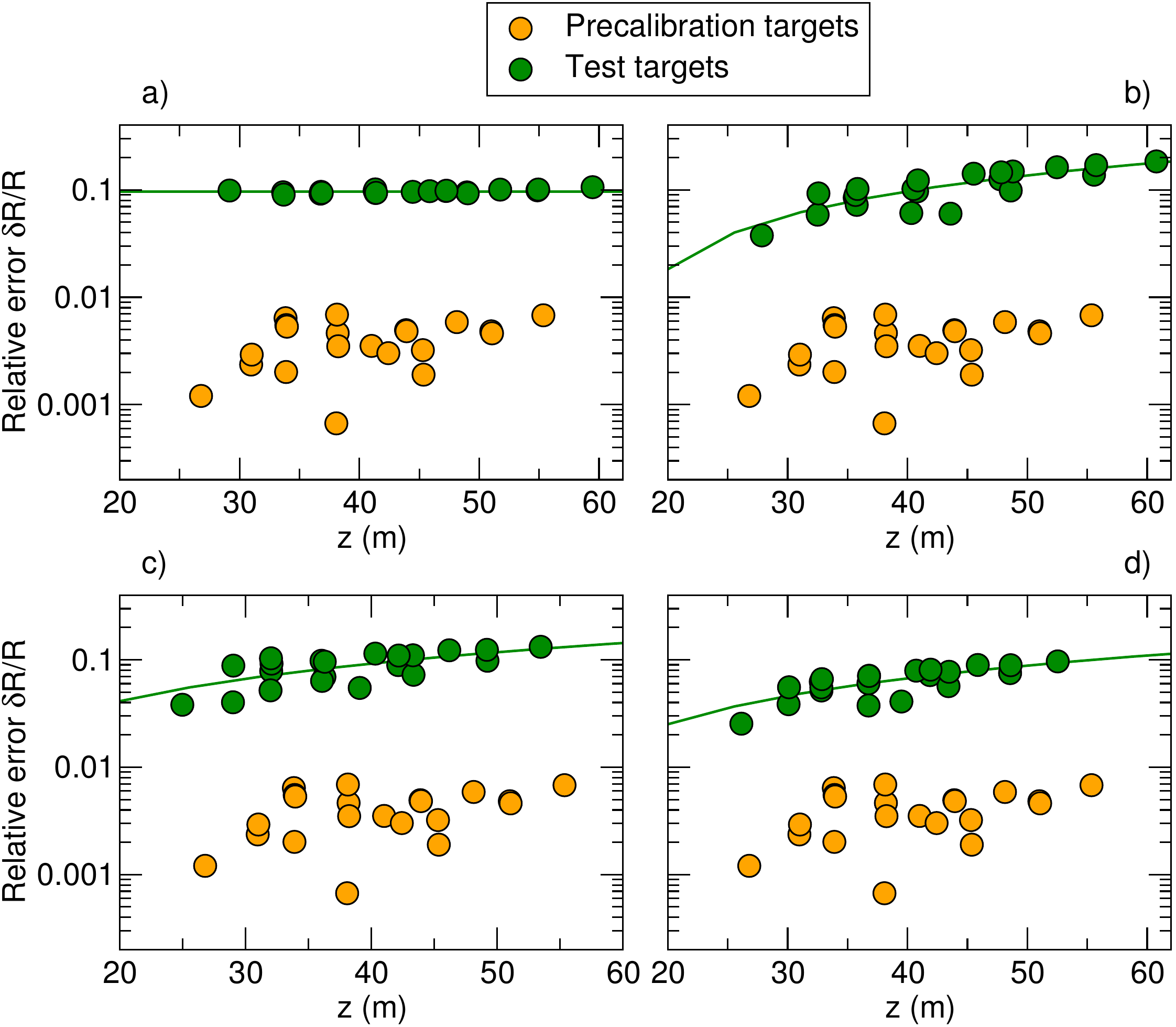}
\caption{{\bf Effects of errors on the measures of intrinsic and extrinsic parameters on the reconstruction of large $\mathbf{R}$.}
{\bf a: error on the measure of the baseline.} Relative reconstruction error  when $\delta d/d=0.1$ on the baseline $d$, green circles, compared with the error using the correct measure of $d$, orange circles. As expected relative errors are constant and equal to $0.1$.
{\bf b: error on the measure of the mutual angle about the y axis.} Relative reconstruction error  when $\delta\alpha=-0.015rad$, green circles, compared with the error using the correct measure of $\alpha$, orange circles. The slope of the linear fit is equal to $0.0039$ corresponding to $2\delta\alpha/d=0.003$. The error is quite big and it reaches the value $0.15$ for $z\sim 60m$.
{\bf c: error on the focal length.} Relative reconstruction error  when $\delta\Omega/\Omega=0.1$, green circles, compared with the error using the correct measure of $\Omega$, orange circles. The slope of the linear fit is equal to $0.0025$ corresponding to $2\alpha\delta\Omega/(\Omega d)=0.003$. Relative error reaches the value $0.15$ for $z\sim 60m$. Note that the term $\Delta z^2\delta\Omega/(R^2\Omega)$ is added to $\delta R/R$, in order to not affect the fit with a quantity not constant for all the pairs of target.
{\bf d: segmentation error.} Relative reconstruction error  when $\delta s=-30px$, green circles, compared with the error using the correct segmentation, orange circles. The slope of the linear fit is equal to $0.0021$ corresponding to $2\delta s/(\Omega d)$. The relative error at $z\sim 60m$ is quite close to $0.1$. 
}
\label{fig:plots}
\end{figure}

The average of the errors on the reconstructed distances is by far the first measure to look in the results of a reconstruction test. But it is also interesting and  more useful to analyze the results looking for sources of errors. The big span of $z$ for targets used in the test, allows a more detailed analysis. Fig.\ref{fig:plots} shows the results on the same reconstruction test of Fig.\ref{fig:birds_precal}, but where we manually added errors on the intrinsic and extrinsic parameters of the system. These results perfectly match the theory described in the paper. 

\begin{figure}[t!]
\includegraphics[width=0.8\columnwidth]{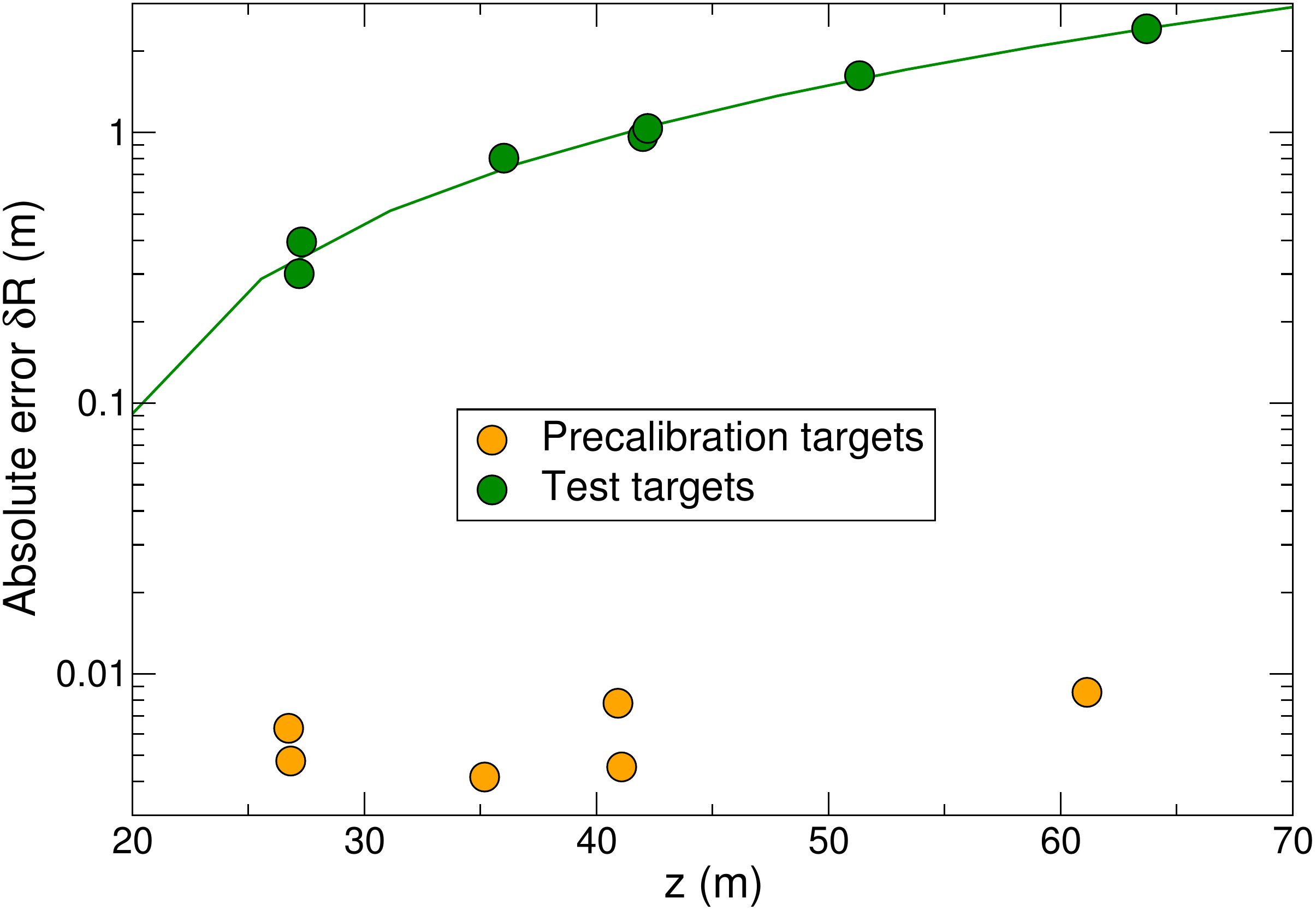}
\caption{{\bf Effect of segmentation error on the reconstruction of short $\mathbf{R}$.}
Relative reconstruction error  when $\delta\Delta s=-10px$, green circles, compared with the error using the correct segmented points, orange circles. The coefficient of the quadratic fit is equal to $0.00047$ and it is compatible with $2\delta\Delta s/(\Omega d)$. 
}
\label{fig:short_distances}
\end{figure}

The constant relative error due to a wrong measure of $d$ is shown in Fig.\ref{fig:birds_precal}a. Fig.\ref{fig:birds_precal}b,\ref{fig:birds_precal}c and \ref{fig:birds_precal}d show the linear trend of the three terms of the relative error on $R$ depending respectively on $\alpha$, $\Omega$ and $s$. Instead in Fig.\ref{fig:short_distances} the effect of a wrong segmentation of targets at short distances of about $0.2m$ are shown. As expected this term is quadratic in $z$ and it reaches $1m$ for $z\sim 50m$.

Note that we forced the system to have errors on intrinsic and extrinsic parameters to be much bigger than the typical experimental errors. A relative error of $0.1$ on $d$ would correspond, in our birds experimental set up, to an absolute error of about $2.5m$, which is not realistic at all. The only reasonable error is the one on $\alpha$, and as shown in Fig.\ref{fig:plots}, it is the one mostly affecting the $3d$ reconstruction accuracy.

Whenever we run a test on the reconstruction quality we plot the relative error on large $\Delta z$ vs $z$; we first look at the average value of the errors. If we obtain high and almost constant errors the most probable cause is a bad measure of $d$ and we check it taking again the distance, or measuring the baseline more carefully. Then we look if there is a linear trend relating the relative error on $\Delta z$ to $z$. If we find a clear linear trend we try to understand if the error is coming from a bad measure of $\alpha$, $\Omega$ or $\delta s$ performing again the test and in the worst case calibrating a new time the intrinsic parameters of the system. 

In the nasty case when we find high reconstruction errors due to a miscalibration of the intrinsic parameters or due to a bad measure of the extrinsic parameters, we throw away the correspondent collected data, so that we are sure that our analysis is based only on reliable trajectories.

\section*{Conclusions}
In the design of a $3d$ experiment the choice of intrinsic and extrinsic parameters is very delicate. A trade off between biological necessity, environmental constraints and accuracy of the reconstruction of the $3d$ position of the imaged targets has to be found. 

In the paper we showed how errors in the measurement of the system parameters affect the reconstruction of the mutual distance between targets. As a consequence they affect the analysis of quantities like velocity, acceleration and correlation functions. Moreover errors on different parameters influence the reconstructed distances depending on their size and on their positions. In particular, large distances are mostly affected by errors on the orientation of the cameras, while short distances by segmentation errors. In the example of Fig.\ref{fig:plots}b, a small error on $\alpha$ of $0.01rad$ produces relative errors up to $0.16$, while in the example of Fig.\ref{fig:short_distances} a segmentation error of $10px$ produces errors up to $1m$ at $z\sim40m$ over mutual distances of about $0.2m$. In our experiment we manage to keep relative errors on large distances smaller than $0.01$ and absolute errors on short distances below $1cm$ (over distances of about $0.2m$). 

Independently on the intrinsic and extrinsic parameters calibration procedures and on the segmentation software used, the best way to reduce the reconstruction error is to design the proper set up. The strategy is to choose large $\Omega$ and $d$ trying to be as close as possible to the group of interest. But at the end of the day, the only way to guarantee the reliability of the retrieved trajectories is to take care of the error while planning the experiment and then test the accuracy.


\begin{thebibliography}{1}
	%

\bibitem{gallup2012} A.C. Gallup, J.J. Hale, D.J.T. Sumpter, S. Garnier, A. Kacelnik, J.R. Krebs, I.D. Couzin, Visual attention and the acquisition of information in human crowds. \emph{PNAS} {\bf 109}, 19, 7245--7250 (2012).

\bibitem{moussaid2009} M. Moussaid, D. Helbing, S. Garnier, A. Johansson, M. Combe, G. Theraulaz, Experimental study of the behavioural mechanisms underlying self-organization in human crowds. \emph{Proc. R. Soc. B} {\bf{276}}, 2755--2762 (2009). 

\bibitem{katz2011} Y. Katz, K. Tunstrom, C.C. Ioannou, C. Huepe, I.D. Couzin, Inferring the structure and dynamics of interactions in schooling fish. \emph{PNAS} {\bf 108}, 46, 18720--18725 (2011).

\bibitem{butail20103d}
	S. Butail, D.A. Paley, 3D reconstruction of fish schooling kinematics from underwater video. \emph{Robotics and Automation (ICRA), 2010 IEEE International Conference on} 2438--2443, (2010).

\bibitem{attanasi2014information}
  A. Attanasi, A. Cavagna, L. Del Castello, I. Giardina, T.S. Grigera, A. Jeli\'{c}, S. Melillo, L. Parisi, O. Pohl, E Shen, and M. Viale, Information transfer and behavioural inertia in starling flocks. \emph{Nature physics} {\bf{10}}, 9, 691--696 (2014).

\bibitem{attanasi2014emergence} A. Attanasi, A. Cavagna, L. Del Castello, I. Giardina, A. Jelic, S. Melillo, L. Parisi, O. Pohl, E. Shen, M. Viale, Emergence of collective changes in travel direction of starling flocks from individual birds fluctuations, arxiv:1410.3330 (2014).

\bibitem{attanasi2014prl}
	A. Attanasi, A. Cavagna, L. Del Castello , I. Giardina, S. Melillo, L. Parisi, O. Pohl, B. Rossaro, E. Shen, E. Silvestri, M. Viale, Finite-size scaling as a way to probe near-criticality in natural swarms. \emph{Phys. Rev. Lett.} {\bf{113}}, 238102, (2014).

\bibitem{attanasi2014collective}
    A. Attanasi, A. Cavagna, L. Del Castello, I. Giardina, S. Melillo, L. Parisi, O. Pohl, B. Rossaro, E. Shen, E. Silvestri, and M. Viale, Collective behaviour without collective order in wild swarms of midges, \emph{PLoS Computational Biology} {\bf{10}}, 7, 1--15 (2014).
    

\bibitem{butail20113d}
	S. Butail, N. Manoukis, M. Diallo, A.S. Yaro, A. Dao, S.F. Traore, J.M. Ribeiro, T. Lehmann, D.A. Paley, 3D tracking of mating events in wild swarms of the malaria mosquito Anopheles gambiae.  \emph{Engineering in Medicine and Biology Society, EMBC, 2011 Annual International Conference of the IEEE} {\bf{75}}, 720--723 (2011).
	
 \bibitem{butail2012reconstructing}
    S. Butail, N. Manoukis, M. Diallo, Ribeiro J. M., T. Lehmann and D. A. Paley, Reconstructing the flight kinematics of swarming and mating in wild mosquitoes. \emph{J. R. Soc. I} {\bf{75}}, 2624--2638 (2012).
    

\bibitem{cavagnaAnimal} A. Cavagna, I. Giardina, A. Orlandi, G. Parisi, A. Procaccini, M. Viale, V. Zdravkovic, The STARFLAG handbook on collective animal behaviour: 1. Empirical methods. \emph{Anim. Behav} {\bf 76}, 217--236 (2008).
	

\bibitem{theriault2014} D. Theriault, N.W. Fuller, B.E. Jackson, E. Bluhm, D. Evangelista, Z. Wu, M. Betke, T.H. Hedrick, A protocol and calibration method for accurate multi-camera field videography. \emph{J. Exp. Biol} {\bf 217}, 1843--1848 (2014).

\bibitem{towne2012} G. Towne, D. H. Theriault, Z. Wu, N. Fuller, T. H. Kunz, and M. Betke, Error Analysis and Design Considerations for Stereo Vision Systems Used to Analyze Animal Behavior. \emph{Proceeding of IEEE Workshop on VAIB}, (2012).

\bibitem{hartley2003book}
	R. Hartley and A. Zisserman, \emph{Multiple View Geometry in Computer Vision}, second ed. Cambridge, U.K.: Cambridge University Press, 2003.

	
\end{thebibliography}
\end{document}